\documentclass[a4paper,11pt]{article}
\pdfoutput=1 
\usepackage{jcappub} 

\hypersetup{
  colorlinks = true,
  linkcolor = red,
  urlcolor  = magenta,
  citecolor = blue,
  anchorcolor = blue
}

\usepackage{subfig}
\usepackage{amsmath, amsfonts,bm}
\usepackage{graphicx}
\usepackage[usenames,dvipsnames]{xcolor}

\DeclareMathOperator{\sech}{sech}
\def\dd{\mathrm{d}}

\newcommand{\be}{\begin{equation}} 
\newcommand{\ee}{\end{equation}}
\newcommand{\ba}{\begin{eqnarray}}
\newcommand{\ea}{\end{eqnarray}}

\newcommand{\lag}[1]{\mathcal{L}_{#1}}
\newcommand{\bA}{{\mathbf{A}}}

\newcommand{\Hd}{\tilde{H}}

\begin{document}

\title{Non-trivial thick brane realisations with 3-forms}

\author[a]{Bruno J. Barros}
\author[b]{Jose Beltr\'an Jim\'enez.}

\affiliation[a]{Instituto de Astrof\'isica e Ci\^encias do Espaço,
Faculdade de Ci\^encias da Universidade de Lisboa,
Campo Grande, PT1749-016 Lisboa, Portugal.}

\affiliation[b]{Departamento de F\'isica Fundamental and IUFFyM, Universidad de Salamanca, E-37008 Salamanca, Spain.}

\emailAdd{bjbarros@fc.ul.pt}
\emailAdd{jose.beltran@usal.es}

\abstract{
We explore the construction of four-dimensional thick branes supported by massless 3-forms in a five-dimensional bulk space. The required residual Poincar\'e symmetry on the brane is realised as a combination of the bulk symmetries and the internal gauge symmetry of the 3-form. We show the presence of a gradient instability for the perturbations orthogonal to the brane when its profile decays slowly in the asymptotic regions. In particular, we ascertain that the prevailing profiles found in the literature are susceptible to such instabilities. We confirm our results by transitioning to the dual formulation in terms of a shift-symmetric scalar. In this formulation, the residual Poincar\'e group is trivially realised without internal symmetries, and there is a diagonal translational invariance in the fifth dimension reminiscent of the invariance under translations orthogonal to the brane of the dual 3-form. We demonstrate the extension of our results to the cases of de Sitter and Anti de Sitter branes where the gradient instabilities of asymptotically slowly decaying branes persist. Finally, we briefly comment on the construction of thick branes with massive 3-forms and their 1-form duals.
}

\date{\today}

\maketitle

\section{Introduction}\label{Sec::intro}

The concept that our Universe could be confined to a brane embedded in a higher dimensional bulk can be traced back to the early 80's \cite{Rubakov:1983bb,10.1007/3-540-11994-9_41}. These studies suggest that particles in the standard model lie within the interior of a vortex or a domain wall, particularly within the framework of topological defects. In these models, the brane is assigned a finite thickness, thus, in contemporary terminology, they are now commonly referred to as thick branes. More than a decade after, two seminal works by Randall and Sundrum \cite{PhysRevLett.83.3370,Randall:1999vf} proposed that our 4-dimensional Universe manifests as a 3-brane immersed in a fifth dimension (bulk) where only gravity propagates. This configuration provides a means to address the hierarchy problem of gravity among the other fundamental forces. Although infinitesimally thin branes are widely discussed in the literature (see e.g., Ref.~\cite{Langlois:2002bb}), from a realistic standpoint, it is more natural to model our four dimensions as a localised brane with a finite thickness.

From a symmetry perspective, thick brane scenarios can be interpreted as 4-dimensional systems confined to a finite region of some higher-dimensional space retaining certain residual symmetries. It is customary to consider Minkowski branes where the residual symmetry corresponds to a 4-dimensional Poincar\'e group. If we have a 5-dimensional Poincar\'e invariance, the brane realises the symmetry breaking pattern ${\text{ISO(4,1)}\rightarrow \text{ISO(3,1)}}$ so that translations and boosts involving the fifth dimension are spontaneously broken. The simplest realisation of this symmetry breaking pattern is with scalar fields that develop a non-trivial profile along the fifth dimension so that the four-dimensional Poincar\'e group orthogonal to its gradient is preserved. For this reason, it stands out as the most widely adopted construction method for thick branes (see \cite{Dzhunushaliev:2009va} and references therein). However, when the matter sector has some internal (global or gauge) symmetries, these can be exploited to realise the desired 4-dimensional Poincar\'e group in non-trivial manners. In these constructions, the translations and the homogeneous Lorentz transformations are realised as the combined action of the spacetime generators together with some of the internal generators. This closely resembles what happens in (flat) cosmological scenarios that can be understood as systems with a residual Euclidean group. In most cases, this residual group coincides with the subgroup of the original Poincar\'e group. However, there are scenarios like gauge-flation \cite{Maleknejad:2011jw}, solid inflation \cite{Endlich:2012pz} or gaugid inflation \cite{Piazza:2017bsd} where the residual ISO(3) group involves some internal transformations. A comprehensive classification of possible non-trivial realisations can be found in \cite{Nicolis:2015sra} (see also \cite{BeltranJimenez:2018ymu}).
These non-trivial realisations also emerge in teleparallel cosmologies \cite{Gomes:2023hyk}, potentially offering an alternative pathway for constructing thick branes. However, we will not delve into this direction. Instead, the goal of this work will be to explore the construction of thick branes supported by massless 3-forms. Since 3-forms do not admit Lorentz invariant background configurations, we will need to resort to their gauge symmetry to construct non-trivial realisations of the brane Poincar\'e invariance.

This work is organized as follows: In Sec.~\ref{sec:brane_sym} we show the explicit construction of the 3-form that realises the brane $\text{ISO(3,1)}$ invariance when acting in combination with a gauge transformation, compute the field equations and discuss the stability of the 3-form when sustaining Minkowskian thick branes. The massless 3-form dualisation process to a shift-symmetric scalar field in five dimensions is presented in Sec.~\ref{sec::scalardual}. In Sec.~\ref{sec:field} we compute the field equations in terms of the dual scalar field, dissect the stability conditions, and, in Sec.~\ref{sec:exmpl}, give an explicit example of a brane solution. Non-flat (de Sitter and Anti de Sitter) thick branes are briefly analysed in Sec.~\ref{sec::nonflat} where we tailor the 3-form for such gravitational framework. Sec.~\ref{sec:Massive} presents a discussion of brane profiles supported by massive 3-forms, and their corresponding dual massive vectors. We then conclude in Sec.~\ref{sec:conclusions}.

\section{Minkowski branes with massless 3-forms}\label{sec:brane_sym}

\subsection{Minkowski branes}
The metric that describes a Minkowski brane can be written as 
\be\label{metric}
\dd s^2=a^2(y)\eta_{\mu\nu}\dd x^\mu\dd x^\nu+\dd y^2\,,
\ee
where $a(y)$ is the warp factor that governs the brane profile, with $y$ denoting the extra dimension. In order to establish a thick brane, this warp factor is usually taken to present a maximum at $y=0$ (or some other position where the brane is localised) and then it quickly decreases. The brane profile may be symmetric if a $Z_2$ symmetry is present, although this is not mandatory as asymmetric branes are also feasible. It is also convenient to parameterise the warp factor as ${a(y)=e^{W(y)}}$ where $W(y)$ is usually referred to as warp function. This introduces a potential source of confusion between $a$ and $W$. However, throughout this work no confusion will arise. In any case, the crucial aspect for this study is that the brane maintains Poincar\'e invariance. This is the symmetry that we will require for the supporting matter fields, specifically a massless 3-form in our case. 

In addition to the requirement of having a Poincar\'e invariant metric on the brane, its profile will also be subject to some conditions in order to have a regular geometry in the 5-dimensional space. Although the condition to have a localised brane is encapsulated in the decaying behaviour of the warp factor in the asymptotic region $\vert y\vert\to \infty$, such a decay cannot be arbitrary. In particular, the regularity of the curvature invariants such as the Kretschmann scalar is guaranteed if we impose that both $W'(y)$ and $W''(y)$ remain finite. For the moment, we will remain agnostic regarding the asymptotic behaviour of the warp factor. However, we will later observe that this behaviour has crucial implications for the stability of the brane.

\subsection{The massless 3-form support}
We then proceed to demonstrate how a massless 3-form $A_{MNP}$ can realise the residual Poincar\'e symmetry in a non-trivial manner.\footnote{That the realisation must be non-trivial is clear since a 3-form cannot develop a Lorentz-invariant background configuration. See Sec. \ref{sec:Massive}.} The massless nature of the 3-form means that the physical quantity will be its field strength, ${H_{ABCD}=4\partial_{[A}A_{BCD]}}$. Hence, the theory will possess a gauge symmetry ${A_{MNP}\to A_{MNP}+\partial_{[M}\theta_{NP]}}$ for an arbitrary 2-form $\theta_{MN}$. A 3-form configuration that fulfils the desired symmetries can be written (up to a gauge transformation) as
\be
A_{\mu\nu\rho}=\alpha\epsilon_{\mu\nu\rho\sigma}x^\sigma\,,
\label{eq:Avev}
\ee
with $\alpha$ being a constant parameter,\footnote{We will see later that this constant in fact coincides with a conserved charge of the system.} and all the other components vanishing. It is possible to absorb the value of $\alpha$ into a rescaling of the brane coordinates, but we will keep it for convenience. In the above expression $\epsilon_{\mu\nu\rho\sigma}$ is the 4-dimensional (Lorentz-invariant) Levi-Civita tensor. It is easy to see that the field strength takes the form
\be
H_{\mu\nu\rho\sigma}=-4\alpha\epsilon_{\mu\nu\rho\sigma}\,,
\label{eq:configurationH}
\ee
while all other components vanish. Since the 3-form is massless, this is the physical quantity and the fact that it is Lorentz invariant shows that it fulfils the required symmetries for the thick brane solutions, i.e., Eq.~\eqref{metric}. At the level of the 3-form, the realisation of Poincar\'e symmetry is understood as a combination of the 4-dimensional spacetime Poincar\'e transformations together with the internal gauge symmetry of the 3-form. Under the simultaneous action of a 4-dimensional translation ${x^\mu\rightarrow x^\mu+x_0^\mu}$ and a gauge transformation ${A_{\mu\nu\rho}\rightarrow A_{\mu\nu\rho}+\partial_{[\mu}\theta_{\nu\rho]}}$ , the configuration \eqref{eq:Avev} changes as
\be
A_{\mu\nu\rho}(x)\rightarrow A_{\mu\nu\rho}+\epsilon_{\mu\nu\rho\sigma} x^\sigma_0+\partial_{[\mu}\theta_{\nu\rho]}\,.
\ee 
This obviously breaks translations on the brane. However, by choosing ${\theta_{\mu\nu}=\alpha\epsilon_{\mu\nu\rho\sigma}x_0^\rho x^\sigma}$ we find that translations are restored. This realises the inhomogeneous part of the brane ISO(3,1) symmetry group. Regarding the homogeneous part, we first note that the 3-form changes under a Lorentz transformation as
\begin{align}
A_{\mu\nu\rho} \;\rightarrow \; \Lambda_\mu{}^{\alpha}\Lambda_\nu{}^{\beta}\Lambda_\rho{}^{\pi}A_{\alpha\beta\pi}\big(\Lambda\cdot x\big)= \;\Lambda_\mu{}^{\alpha}\Lambda_\nu{}^{\beta}\Lambda_\rho{}^{\pi}\epsilon_{\alpha\beta\pi\sigma}\Lambda^{\sigma}{}_\delta x^\delta\,.
\end{align}
In order to show the residual homogeneous Lorentz symmetry, it will be more convenient to consider infinitesimal transformations, i.e., ${\Lambda^\mu{}_\nu\simeq\delta^\mu_\nu+\omega^\mu{}_\nu}$ with $\omega^\mu{}_\nu$ the Lorentz generators that satisfy ${\omega_{\mu\nu}=-\omega_{\nu\mu}}$. Considering only proper transformations, we have $\det(\Lambda)=1$ and the following relation for the generators
\be
\omega^\sigma{}_\pi\epsilon_{\alpha\beta\gamma\sigma}=\omega_\alpha{}^\sigma\epsilon_{\sigma\beta\gamma\pi}+\omega_\beta{}^\sigma\epsilon_{\alpha\sigma\gamma\pi}+\omega_\gamma{}^\sigma\epsilon_{\alpha\beta\sigma\pi}\,.
\ee
Under the infinitesimal transformation, the 3-form changes as
\be
\delta A_{\mu\nu\rho} = 2\alpha\epsilon_{\mu\nu\rho\pi}\omega^\pi{}_\sigma x^\sigma\,.
\ee
Again, we corroborate that a Lorentz transformation alone does not preserve the background configuration Eq.~\eqref{eq:Avev}.
However, we can compensate it by simultaneously taking advantage of the internal symmetry of the field. Let us consider the gauge transformation with
\be
\theta_{\nu\rho}=\frac\alpha2\epsilon_{\nu\rho\alpha\beta}\Omega^{\alpha\beta}x^2\,,
\ee
where $\Omega^{\alpha\beta}$ is some constant and antisymmetric parameter. Under this transformation, the 3-form potential changes as
\be
\delta A_{\mu\nu\rho}=\partial_{[\mu}\theta_{\nu\rho]} = \alpha x_{[\mu}\epsilon_{\nu\rho]\alpha\beta}\Omega^{\alpha\beta}=-\frac{2\alpha}{3}\epsilon_{\mu\nu\rho\sigma}\Omega^\sigma{}_\lambda x^\lambda\,.
\ee
If we choose ${\Omega^{\alpha\beta}=3\omega^{\alpha\beta}}$, this gauge transformation cancels the change induced by the Lorentz transformation, thus ensuring that the 3-form configuration remains invariant.\footnote{Let us note that the gauge transformation itself has a gauge symmetry since $\theta_{\mu\nu}$ and ${\theta_{\mu\nu}+\partial_{[\mu}\theta_{\nu]}}$ generate the same gauge transformation for the 3-form. This means that the gauge transformation that restores the Lorentz invariance of the background is not unique. For instance, the parameter ${\theta_{\nu\rho} = -3\alpha\epsilon_{\nu\rho\alpha\beta}\omega^\alpha{}_\sigma x^\sigma x^{\beta}}$ would be equally valid.} This is the residual Lorentz symmetry on the brane. 

It is crucial to note that everything mentioned thus far does not depend on any specific action. The realisation of the residual Lorentz symmetry occurs off-shell, with the only requirement being that the 3-form is massless and, consequently, possesses a gauge symmetry. Let us now proceed into the examination of the underlying dynamics.

\subsection{Brane equations}\label{sec::brane_eqs}

The massless nature of the 3-form requires the Lagrangian to depend solely on its field strength. Therefore, configuration \eqref{eq:configurationH} will be a trivial solution of the equations. This can be straightforwardly checked by considering a 3-form Lagrangian 
\be
\lag{A}=F(Y)\,,
\ee
with ${Y=-\frac{1}{48} H_{ABCD} H^{ABCD}}$. The equations derived from this Lagrangian are
\be
\partial_A\left(\sqrt{-g}F_YH^{ABCD}\right)=0\,,
\ee
with ${F_Y=\dd F/\dd Y}$.\footnote{We will never use $Y$ and $X$ as five-dimensional indices so no confusion will arise by denoting the derivative with a subscript.} Since the field strength, Eq.~\eqref{eq:configurationH}, has vanishing components in the $y-$th direction and the warp factor of the metric only depends on such a coordinate, the equation is trivially satisfied for the constant field strength of our background configuration. One could object this line of reasoning if the considered Lagrangian is sufficiently general or if there are other independent gauge-invariant Lorentz scalars that could enter as an argument of $F$. It turns out that, at lowest order in derivatives, the considered Lagrangian is the most general one that can be constructed. On the other hand, the considered Lagrangian describes a consistent effective field theory for the 3-form even in regimes where the interactions in $F$ are non-perturbative. The reason being that quantum corrections will enter with additional derivatives acting on $H_{MNPQ}$. If $M$ is the scale that controls the non-linearities in $F$, it is possible to have a regime where ${H/M^2\gtrsim 1}$ while ${\partial H\ll M H}$.

Another remarkable property of the 3-form configuration is that it does not depend on the fifth dimension, thus exhibiting a translation invariance in the direction orthogonal to the brane. A fair question would then be what determines the position of the brane. This has an obvious answer since, although the 3-form configuration does not depend on $y$, the metric does. In turn, this makes the physical quantities like the energy density or the pressures/stresses of the 3-form to depend on $y$ as we will show in the following. In fact, the setting under consideration leads to ${Y=8\alpha^2 e^{-8W}}$, which does depend on $y$ through the warp function $W$. It is also useful to compute the dual of the field strength 
\be
\Hd^A=\frac{1}{4!}\epsilon^{ABCDE}H_{BCDE}\,,
\ee
that is given by ${\Hd^A=4\alpha e^{-4W}\delta^A_y}$ and we can alternatively write ${Y=\Hd^2/2}$. The energy-momentum tensor for the 3-form is given by 
\be
T_{MN}=\frac16 F_Y H_{MABC} H_N{}^{ABC}+g_{MN}F\,,
\ee
or, in terms of the dual
\be
T_{MN}=F_Y \Hd_{M} \Hd_N+\Big(F-2YF_Y\Big)g_{MN}\,.
\ee
The energy density of the 3-form can then be identified as
\be
\rho=2Y F_Y-F\,,
\ee
while the pressures on the brane and in the orthogonal direction are
\be
p_{\text{brane}}=-\rho,\quad p_y=F\,.
\ee
These expressions show that indeed the physical quantities do depend on the coordinate $y$ and this will fix the mean position of the brane.

The Einstein equations can be found by varying the total action with respect to the metric, and read
\ba
3\left(W''+2W'^2\right) &=& F-2Y F_Y\,,\label{field1A} \\
6W'^2 &=& F\label{field2A}\,,
\ea
where a prime denotes a derivative with respect to the fifth dimension, $y$. These equations formally resemble the usual cosmological equations with Hubble function $W'$, where the roles of the energy density and the pressure are exchanged and $y$ plays the role of cosmological time. Due to this cosmological analogy and inspired by the description of inflationary cosmological solutions as quasi-de Sitter phases described by slow roll parameters, we will describe the asymptotically decaying profile of the brane in terms of
\be
\varepsilon_1\equiv-\frac{W''}{W'^2}=\frac{4YF_Y}{F}=4\frac{\dd \log F}{\dd\log Y}\,,
\ee
that we will call first slow warping parameter. This parameter describes how fast the warp factor decreases. In particular, we have the relation
\be
\frac{a''}{a}=(W')^2(1-\varepsilon_1)\,,
\ee
that shows how a brane with a finite width and ${a''>0}$ far from its mean position demands ${\varepsilon_1<1}$.  The condition ${\varepsilon_1\ll1}$ requires that the Lagrangian $F$ of the 3-form is sufficiently flat, i.e., we need an approximate scale invariance.

We can also introduce the second slow warping parameter
\be
\varepsilon_{2}\equiv\frac{\dd \log\varepsilon_1}{W'\dd y}
\ee
to describe how fast the derivative of the brane profile decays. We can use that
\be
\frac{\dd Y}{\dd y}=-8YW'
\ee
to express the second slow warping parameter in the more useful form
\be
\varepsilon_{2}=-8\frac{\dd \log\varepsilon_1}{\dd Y}=-8\left(1-\frac{YF_Y}{F}+\frac{YF_{YY}}{F_Y}\right)= 2\varepsilon_1-8\left( 1+\frac{YF_{YY}}{F_Y} \right)\,.
\ee
It is not difficult to see that these two parameters capture the slowly decaying brane profiles of most constructions in the literature. In particular, very much like the inflationary slow roll parameters quantify the deviations from de Sitter, the introduced slow warping parameeters can be used to describe deviations from asymptotically maximally symmetric spaces. Our formalism could thus be used to design thick brane scenarios in a systematic way and analyse their dynamics. We are not interested in performing a general dynamical analysis of the construction of these scenarios, but our aim is instead to diagnose their generic stability.

\subsection{Stability}
\label{sec::brane_stab}

After obtaining the thick brane equations supported by a massless 3-form, we will study the stability of the brane construction under perturbations. In general, we should include all possible perturbations, i.e., perturbations of the 3-form together with metric perturbations. However, we will be interested in high frequency perturbations of the 3-form and this will be sufficient for us to unveil a UV instability. Being a UV instability, the coupling to gravity cannot cure it so neglecting them in our analysis will invalidate our results. Thus, let us consider perturbations of the 3-form around the background Eq.~\eqref{eq:Avev}. It is most convenient to use the field strength dual perturbations $\delta \tilde{H}^M$, in terms of which the quadratic Lagrangian reads
\be
\lag{}^{(2)}=\frac{1}{2}\left(F_Yg_{MN}+F_{YY}\tilde{H}_M\tilde{H}_N\right) \delta \tilde{H}^M\delta \tilde{H}^N\,.
\ee
From this quadratic Lagrangian, we can already obtain stability conditions without much effort by noticing that the perturbations of the 3-form (in the considered regime) propagate on an effective metric that depends on the background configuration. As expected, we need to impose $F_Y>0$ to avoid ghosts. This condition already imposes some constraints on the brane profile. By subtracting Eqs.~\eqref{field1A} and \eqref{field2A} we can obtain
\be
3W''=-2YF_Y\,.
\label{eq:WppA}
\ee
Since $Y>0$, the absence of ghosts requires the brane profile to satisfy ${W''<0}$, which is consistent with the requirement for the brane profile to have a maximum. Interestingly, the above expression also tells us that the brane profile cannot exhibit a minimum since that would require ${W'=0}$ and ${W''>0}$ that would be incompatible with the absence of ghosts. Let us note that the right-hand side of Eq.~\eqref{eq:WppA} precisely determines whether the null energy condition will be violated. For our configuration, the null energy condition ${T_{MN}n^M n^N>0}$ for all null vectors $n^M$ reads ${2YF_Y(n^y)^2}$. For a null vector on the brane, this condition saturates to the one of a cosmological constant on the brane ${\rho+p_{\text{brane}}=0}$, while for a null vector with a component normal to the brane the null energy condition reads 
\be
\rho+p_y=2YF_Y>0\,. 
\ee
Thus, we obtain that indeed a brane profile with ${W''<0}$ is consistent with the null energy condition and in turn having ${W''>0}$ requires a violation of such a condition with the associated pathologies that this brings about \cite{Dubovsky:2005xd}. This is equivalent to the outcome wherein a bouncing cosmology or a wormhole solution requires the violation of the null energy condition.

Another stability condition that we need to impose is the requirement for the effective metric to be Lorentzian. It is not difficult to see that the effective metric has three different eigenvalues given by
\begin{align}
\lambda_0=-F_Y\,,\quad\lambda_{\text{brane}}=F_Y\, ,\quad
\lambda_y=F_Y \left(1+\frac{2YF_{YY}}{F_Y}\right)\,,
\end{align}
where $\lambda_0$ and $\lambda_{\text{brane}}$ give the Lorentzian signature of the metric on the brane, while $\lambda_y$ is associated to the effective metric eigenvalue in the extra dimension. Thus, for the effective metric to be Lorentzian, and using that $F_Y$ must be positive to avoid ghosts, we need to have
\be
1+\frac{2YF_{YY}}{F_Y}>0\,.
\label{eq:stabilityA}
\ee
This condition simply reflects the fact that the propagation speed of the perturbations that propagate in the $y$-direction should be real (i.e., the propagations speed squared needs to be positive). The 3-form only contains one propagating degree of freedom and its perturbations are adiabatic.\footnote{All this will be confirmed in Sec. \ref{sec::scalardual} where we will obtain its dual formulation in terms of a scalar field.} This means that the pressure and density perturbations will be proportional, being the proportionality constant the propagation speed squared. For perturbations on the brane, the propagation speed is 1 because the background profile is trivial there and the brane is Minkowskian. On the other hand, the propagation speed of the perturbations along the extra-dimension can be computed as
\be
c_y^2=\frac{\dd p_y}{\dd\rho}=\frac{\dd p_y/\dd Y}{\dd \rho/\dd Y}=\frac{1}{1+2\frac{Y F_{YY}}{F_Y}}\,.
\label{eq:csy2B}
\ee
Thus, as anticipated above, the condition \eqref{eq:stabilityA} guarantees the absence of gradient instabilities that would be triggered by having ${c_y^2<0}$. We can now use Eq.~\eqref{eq:csy2B} to express the propagation speed in terms of the slow warping parameters as
\be\label{cs2}
c_y^2=-\frac{1}{1-\frac12 \varepsilon_1+\frac14 \varepsilon_2}\,.
\ee
This relation tells us that a slowly decaying brane profile such that $\varepsilon_{1,2}\ll 1$ will necessary suffer from gradient instabilities for the perturbations propagating along the brane profile direction. This result is analogous to the gradient instabilities of cosmological quasi-de Sitter solutions with 2-forms \cite{Aoki:2022ylc}, which is reminiscent of what occurs in shift-symmetric scalar field theories (see e.g. \cite{Finelli:2018upr}). This relation to shift-symmetric scalars will also be unveiled for our scenario in the next Section. Before that, let us illustrate the discussed non-trivial constraints on the brane profile by considering some common asymptotic brane behaviours.\footnote{In the following, we will focus on the asymptotic region $y\to\infty$, but our arguments will be equally valid for $y\to-\infty$.} 

Firstly, let us notice that a warp factor that decays as a power law $e^{W}\simeq (y_0/y)^p$ with $p>0$ is not possible because it has $W''\simeq p/y^2$ which is incompatible with the ghost-free condition. In particular, this excludes brane profiles that are analytic at infinity. Thus, as a less obvious case,
let us consider a brane that asymptotically behaves as $e^{W}\simeq e^{-(y/y_0)^p}$ for some positive $y_0$ and $p$. The second derivative of the warp factor is now ${W''\simeq -p(p-1)\left(\frac{y}{y_0}\right)^p y^{-2}}$ and the non-ghost condition imposes that $p$ must be strictly bigger than 1. This excludes common brane profiles in the literature that decay exponentially as $e^{-y/y_0}$ far from the brane position. We further have
\be
\varepsilon_1\simeq\frac{p-1}{p}\left(\frac{y}{y_0}\right)^{-p}\quad\text{and}\qquad \varepsilon_2\simeq\left(\frac{y}{y_0}\right)^{-p}\,.
\ee
In the asymptotic region $y\gg y_0$, both of these parameters are negligible and we explicitly see how this behaviour will be prone to a gradient instability.

\section{The scalar dual}\label{sec::scalardual}

The brane construction that we have discussed so far is evidently nothing but the dual description of a thick brane supported by a shift-symmetric scalar field. In this section, we will show this duality scheme and re-obtain the viability conditions in terms of the dual formulation.

\subsection{Dualisation}

In order to show the duality, let us start from the first order formulation of a theory for a scalar field $\phi$ with a shift-symmetry described by
\be\label{lag_phi}
\lag{\phi}=\pi^A\partial_A\phi-F(\pi^2)\,.
\ee
We can integrate out the auxiliary field $\pi^A$ by solving its algebraic equation of motion
\be\label{eom_pi}
\partial_A\phi=2\frac{\partial F}{\partial \pi^2}\pi_A\,,
\ee
that allows to express $\pi^A$ as a function of $\phi$ and $\partial_A\phi$. More explicitly, by squaring the above equation, solving for $\pi^2$ as a function of $(\partial\phi)^2$ and plugging the resulting on-shell relation back into the Lagrangian, we arrive at the second order formulation
\be
\lag{\phi}=P(X)\,,
\ee
with ${2X =- \partial^A\phi\partial_A\phi}$ and the Lagrangian is given by
\be
P(X)\equiv\left[\pi^A\partial_A\phi-F(\pi^2)\right]_{\pi_A(\partial_A\phi)}\,.
\ee
On the other hand, we can, instead, obtain the equation for $\phi$,
\be\label{eq_phi}
\partial_A\pi^A=0\,,
\ee
that is solved by 
\be\label{pi_vector}
\pi^A=\epsilon^{ABCDE}\partial_{[B} A_{CDE]}\,,
\ee
with $\bA$ some 3-form. By squaring this relation we obtain ${\pi^2=-24(\partial_{[B} A_{CDE]})^2}$ that we can plug back into the Lagrangian to show the equivalence with the Lagrangian
\be\label{L_massless_3f}
\lag{A}=F(H^2)\,,
\ee
where $H_{ABCD}$ is the field strength of the 3-form, {\bf H=dA}. Naturally, Eq.~\eqref{pi_vector} identifies the $\pi$ field as the dual vector of {\bf H}, i.e., {\bf ${\pi=\star H}$}, and Eq.~\eqref{eq_phi} expresses the fact that {\bf H} is a closed form, {\bf dH=0}. Thus, we identify the equivalence between the shift-symmetric scalar field and the massless 3-form in 5 dimensions. From this relation, it is easy to understand how the configuration \eqref{eq:Avev} provides the dual description of a thick brane supported by a shift-symmetric scalar with profile ${\phi=\phi(y)}$. We can also start from a 3-form and perform the inverse dualisation to rewrite it in terms of a shift-symmetric scalar field.

In Appendix \ref{appendix} we provide a detailed discussion and derivation of all the relations between the dual descriptions. Here we will only need the relations for the dual Lagrangians  that hold on-shell:
\ba
P &=& F - 2Y F_Y\,,
\label{dual1} \\
F &=& P-2XP_X\,,
\label{dual2}
\ea
as well as
\be
F_Y + 2YF_{YY}= \frac{1}{P_X + 2XP_{XX}}\,.
\ee
We also have the following useful relations 
\be
P_X=\frac{1}{F_Y}\,\quad\text{and}\quad XP_X=-YF_Y.
\ee
After having shown the duality, it will be instructive to re-obtain the results for the thick brane supported by a 3-form from the dual perspective of a shift-symmetric scalar field.

\subsection{Scalar dual brane equations}\label{sec:field}

The dynamical equation for the shift-symmetric scalar field is given by
\be
\partial_A\Big(\sqrt{-g}P_X\partial^A\phi\Big)=0\,.
\ee
This equation has the form of a conservation law ${\partial_A J^A}$ due to the shift-symmetry ${\delta\phi=c}$ of the theory, where $J^A$ is the conserved Noether current. For the thick brane metric~\eqref{metric}, the field profile compatible with the symmetries is simply $\phi(y)$. This of course realises the brane Poincar\'e symmetry in a trivial manner. We can unveil the translational invariance in the direction orthogonal to the brane by using the scalar field profile as the $y$ coordinate,\footnote{Given that the profile is not monotonic, it may be necessary to select different patches to cover the entire brane. However, this is not pertinent to our context.} so that ${\phi(y)=y}$. Then, if we simultaneously perform a translation ${y\to y+y_0}$ and an internal shift ${\phi\to\phi+c}$ we have ${\delta\phi=y+y_0+c}$ so we have a diagonal translational invariance upon the choice ${c=-y_0}$. Thus, we find that the brane symmetries are trivially realised in the scalar field perturbation, but the invariance of the field configuration under translations in the fifth-dimensions is non-trivially realised, while in the 3-form description is the other way around.

The profile of the scalar field $\phi(y)$ on the metric~\eqref{metric} is then determined by the equation
\be
\partial_y\Big(e^{4W}P_X\phi'\Big)=0\,,
\ee
that can be directly integrated to give
\be
e^{-4W}P_X\phi'=C\,,
\ee
where $C$ is the conserved charge associated to the shift-symmetry of the scalar field. 
Since far away from the position of the brane the profile is such that ${e^{2W}\ll1}$, we have that the scalar field must behave so that ${P_X\phi'\gg1}$. This can also be expressed as ${-2XP_X^2=C^2e^{-8W}}$. Since we have the on-shell relation ${XP_X^2=-2Y}$ (see Eq.~\eqref{aux}), we can obtain ${2Y=C^2e^{-8W}}$. Recalling that ${Y=8\alpha^2 e^{-8W}}$, we can identify the conserved charge of the scalar field as ${C=4\alpha}$, thus confirming that $\alpha$ is proportional to a conserved charge.

\begin{figure*}[t!]
      \subfloat{\includegraphics[height=0.325\linewidth]{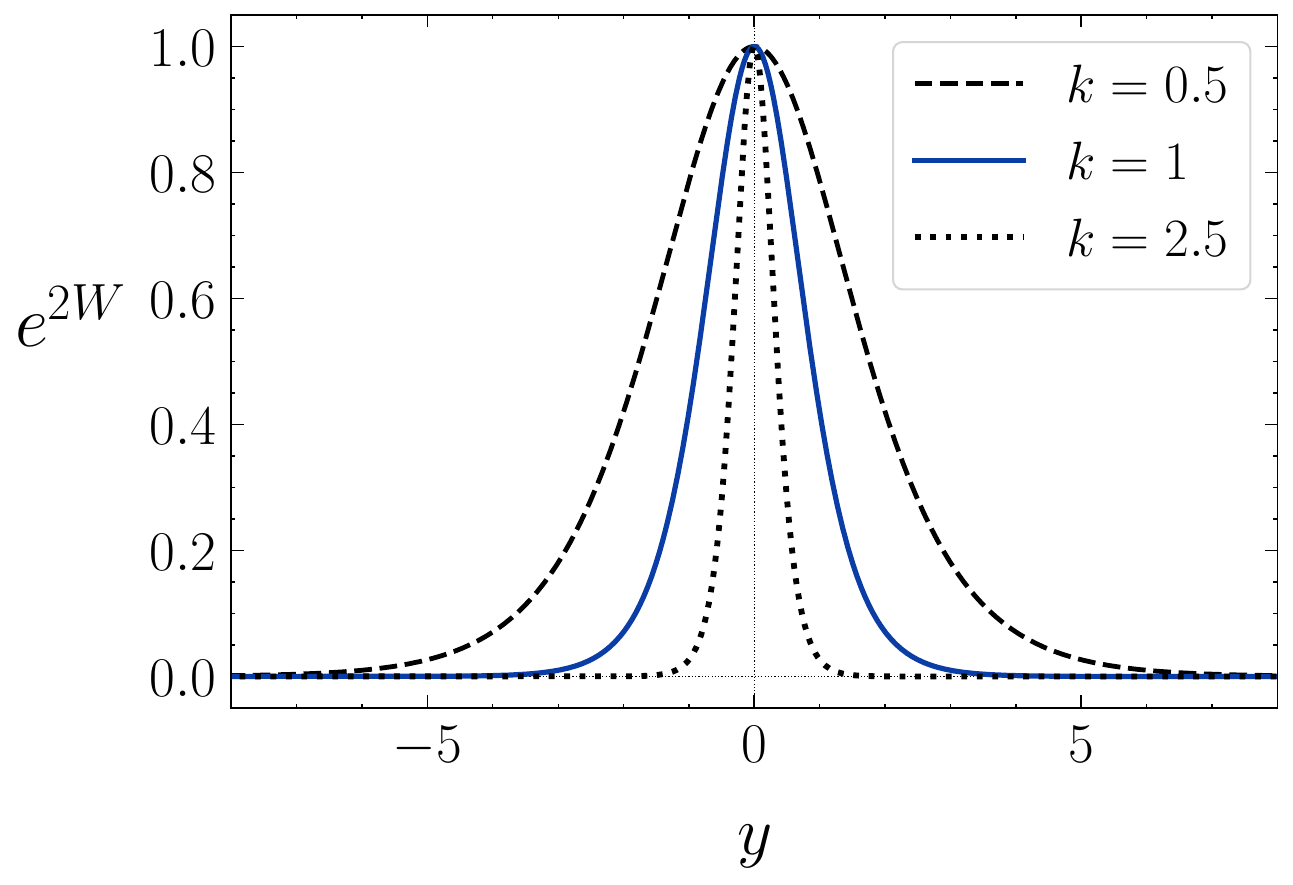}}
      \qquad
      \subfloat{\includegraphics[height=0.325\linewidth]{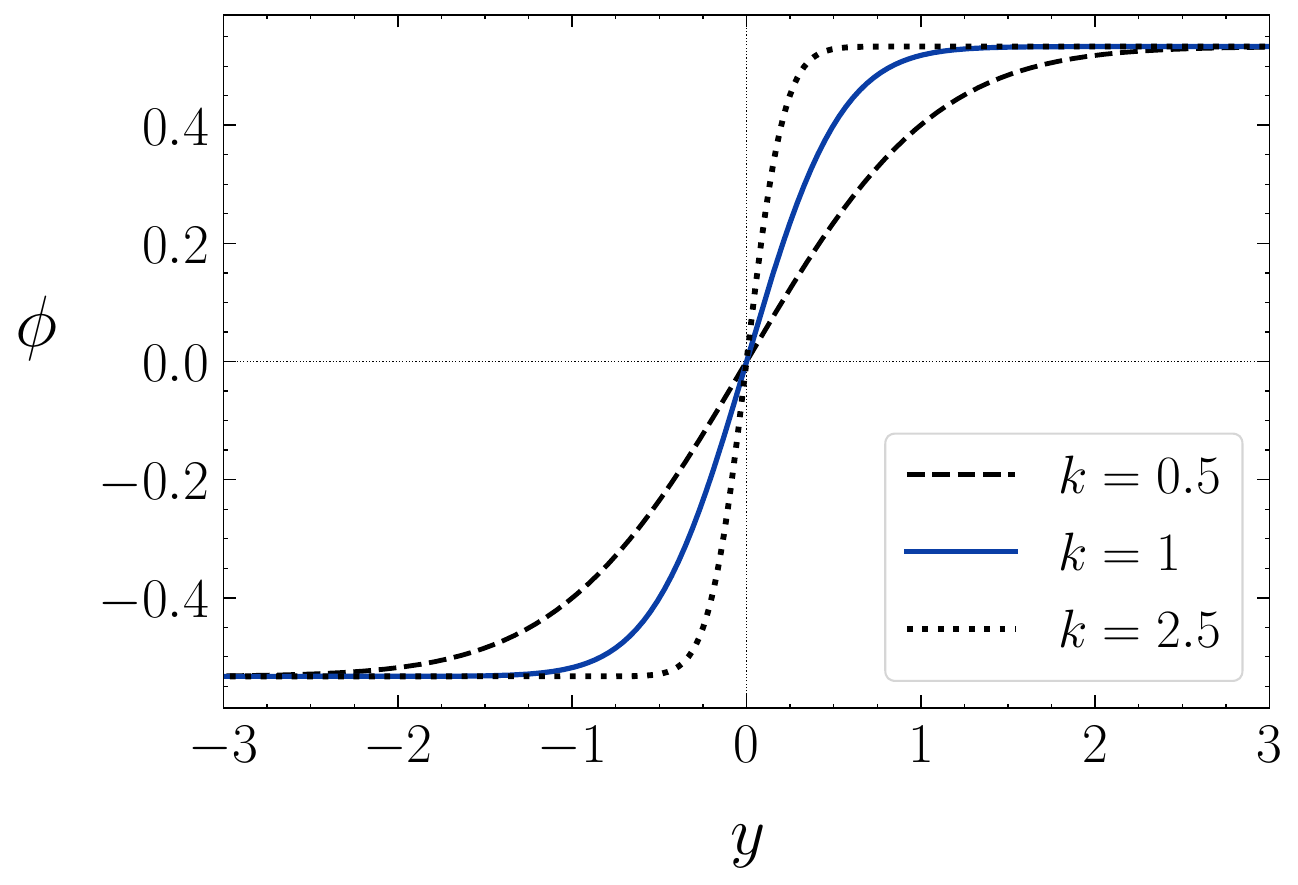}}
  \caption{\label{fig:ex1}Left panel: warp factor, $-g_{00}$, with the choice given by Eq.~\eqref{w_func_ansatz} for different values of $k$. Right panel: field profile given by Eq.~\eqref{phi_sol} with $A=0$, $B=1$ and different values of $k$.}
\end{figure*}

Let us now turn to the gravitational equations. The energy-momentum tensor is
\be
T_{MN}=P_X\partial_M\phi\partial_N\phi+Pg_{MN}
\ee
from which we identify the energy density and the pressures as
\be
\rho=-P,\quad p_{\text{brane}}=P\quad\text{and}\quad p_y=P-2XP_X\,.
\ee
The Einstein equations read
\ba
3\left(W''+2W'^2\right) &=& P\,,\label{field1} \\
6W'^2 &=& P-2XP_X\,.\label{field2}
\ea
These equations exactly coincide with Eqs.~\eqref{field1A} and \eqref{field2A} upon use of the duality relations.

We can compute the slow warping parameters, introduced in Sec.~\ref{sec::brane_eqs}, in terms of the scalar field:
\begin{align}
\varepsilon_1&=-\frac{4XP_X}{P-2XP_X}\,,\\
\varepsilon_2&=-8\frac{P_X+XP_{XX}}{P_X+2XP_{XX}}+2\varepsilon_1\,,
\label{eq:epsilon2phi}
\end{align}
which again coincide with the parameters obtained in terms of the 3-form when using the duality relations. These parameters are small if ${XP_X/P\ll1}$ and ${XP_{XX}/P_X\ll1}$, which are the dual conditions for the brane to have a slow decay in the asymptotic region far from the brane. 

\subsection{Stability}\label{sec::stability}

We will now proceed to studying the perturbations from the dual perspective of the scalar field. Again, we will only consider perturbations of the scalar field that will be sufficient for studying the stability against high frequency modes. The quadratic Lagrangian for the scalar field perturbations ${\delta\phi(x^\mu,y)}$ is given by
\be
\lag{}^{(2)}=-\frac12\left[P_X\eta^{\mu\nu}\partial_\mu\delta\phi\partial_\nu\delta\phi+\big(P_X+2XP_{XX}\big)(\partial_y\delta\phi)^2\right]\,.
\ee
This quadratic action yields the usual constraint ${P_X>0}$ to avoid ghosts, which is the dual of ${F_Y>0}$. As in the 3-form description, this non-ghost condition gives constraints on the brane profile. Combining the gravitational Eqs.~\eqref{field1} and \eqref{field2}, we can obtain the relation
\be\label{field_combined}
3W''=2XP_X\,.
\ee
Noting that ${X<0}$, the equation above once again illustrates how having ${W''<0}$ at the mean position of the brane is consistent with the ghost-free condition. This is guaranteed if the null energy condition, ${\rho+p_y=-2XP_X}$, holds. 

From the quadratic action for the scalar field, we confirm that the perturbations on the brane propagate at the speed of light, while the propagation speed of the perturbations orthogonal to the brane propagate at the speed
\be
c_y^2=1+\frac{2XP_{XX}}{P_X}\,.
\ee
This propagation speed can be seen to coincide with the adiabatic speed $c_y^2=\frac{\dd p_y}{\dd \rho}=\frac{p_X}{\rho_X}$ as we argued above for the 3-form. By using the duality relations, we can also see that this is the same propagation speed as we obtained for the 3-form. This equality can be shown explicitly by using Eq.~\eqref{eq:epsilon2phi} to express the propagation speed as
\be
c_y^2=-\frac{1}{1-\frac12\varepsilon_1+\frac14\varepsilon_2}\,,
\ee
which is the same expression obtained for the 3-form. We thus confirm that the regime where $\varepsilon_{1,2}\ll1$ comes hand in hand with the presence of Laplacian instabilities for the perturbations in the extra dimension. Thus, scalar field scenarios in which the kinetic energy dominates the asymptotic regions (e.g., Ref.~\cite{Bazeia:2008zx}) will be prone to this pathology.

\subsection{An explicit example}\label{sec:exmpl}

As aforementioned in Sec.~\ref{sec:field}, there are two independent equations that completely describe the background of the thick brane. Since we have a total of three independent variables - one geometrical (warp) function, $W$, the field, $\phi$, and the Lagrangian, $P$ - we have the freedom to assume a specific form for one of these. In order to find a specific solution describing a localised brane, let us then consider the following ansatz for the warp function, well represented in the literature \cite{Koley:2004au,Bogdanos:2006qw,Slatyer:2006un,Rosa:2021tei}:
\be\label{w_func_ansatz}
W(y)=\log\left[\sech\left(ky\right)\right]\,,
\ee
which gives ${g_{00}=\sech^2(ky)}$, with $k$ being a constant related to the width of the brane. The demeanour of the brane configuration is depicted in the left panel of Fig.~\ref{fig:ex1}, where we plot the warp factor, ${e^{2W}}$, for different values of $k$. With this framework at hand, it is straightforward to find the specific scalar field Lagrangian which reproduces the geometry given by Eq.~\eqref{w_func_ansatz}.

From Eq.~\eqref{field1} one finds for the scalar Lagrangian,
\be\label{P_func_sol}
P=3k^2\left[2-3\sech^2\left( ky\right)\right]\,.
\ee

We can now plug in the above result to Eq.~\eqref{field2} to find a second order equation for the scalar field,
\be
\phi'' +6k\phi'\tanh\left( ky \right)=0\quad{\text{or}}\quad \partial_y\log\phi' = -6k\tanh\left(ky\right)\,,
\ee
which has the following analytical solution
\be\label{phi_sol}
\phi = A+B\left[ \tanh\left(ky\right)-\frac{2}{3}\tanh^3\left(ky\right) + \frac{1}{5}\tanh^5\left(ky\right) \right]\,,
\ee
with both $A$ and $B$ constants. The dynamics of the field are depicted in the right panel of Fig.~\ref{fig:ex1}. Both the parameters $A$ and $B$ have no influence whatsoever on the brane geometry, and just set the amplitude of the field transition,
\be
\lim_{y\rightarrow\pm\infty}\phi(y) = A\pm \frac{8}{15}B\,.
\ee
On the other hand, $k$ controls the thickness of the brane: higher values of $k$ result on a faster field transition and a thinner brane profile.

Taking the derivative of the field profile, Eq.~\eqref{phi_sol}, one finds,
\be
X = -\frac{1}{2}\phi'^2 = -\frac{B^2}{2}\sech^{12}\left(ky\right)\,,
\ee
which substituting into Eq.~\eqref{P_func_sol} shows the explicit $P$ dependence on $X$, i.e.,
\be\label{lagrangian_final}
P(X)=6k^2\left[1-\frac{3}{2}\left(-\frac{2X}{B^2}\right)^{1/6}\right]\,.
\ee

Making use of the dualization relations, in particular Eq.~\eqref{dual31}, gives ${Y\propto (-X)^{-3/2}}$, which plugging into the expression for $F$ in Eq.~\eqref{dual2}, we find the explicit expression for the 3-form dual theory,
\be
F(Y) = 6k^2\left[ 1-\frac{\sqrt{3}k}{\left( 2B^2Y \right)^{1/4}} \right]\,.
\ee
Using Eq.~\eqref{cs2} we find a constant speed of propagation along the brane of ${c_y^2=-2/3}$. This is in agreement with the results found on the previous section in regard to the presence of Laplacian instabilities. As a final note, a curious solution can be found by expanding the warp function, Eq.~\eqref{w_func_ansatz} around ${y=0}$,
\be \log\left[\sech\left(ky\right)\right] \approx -\frac{k^2y^2}{2}+\mathcal{O}\left(y^3\right)\,, \ee
which, naturally, obeys the geometrical properties required to represent a thick brane (pathological nonetheless). By assuming the above warp factor a priori one would obtain a solution with,
\be
P(X)\propto \log (-X) \Rightarrow F(Y)\propto \log Y \,,
\ee
with ${X\propto\exp (-4k^2y^2)}$ and ${c_y^2=-1}$.

\section{Non-Flat thick branes}\label{sec::nonflat}

Although in the previous sections we have focused on flat branes, the generalisation to non-flat branes is straightforward. To do so, we simply need to consider a different symmetry breaking pattern where the residual symmetry group on the brane is either SO(4,1) or SO(3,2). The metric can be written as
\be
\dd s^2=e^{2W(y)} \gamma_{\mu\nu}\dd x^\mu\dd x^\nu+\dd y^2
\ee
where $\gamma_{\mu\nu}$ is the brane metric with constant curvature ${R=4\Lambda}$. Thus, we have de Sitter, Anti de Sitter and Minkowski for ${\Lambda>0}$, ${\Lambda<0}$ and ${\Lambda=0}$, respectively. Let us notice that these three maximally symmetric geometries on the brane can be obtained as the little groups of an SO(4,2) symmetry group corresponding to space-like, time-like and null vectors. Furthermore, the flat brane can be obtained by means of an Inonu-Wigner contraction \cite{inonuwigner} of the non-flat branes. We will not explore these interesting geometrical relations here, since our goal is to simply show how our construction for the Minkowski branes can be straightforwardly extended. The non-flat branes will be characterised by a 3-form whose field strength is again of the form
\be
H_{\mu\nu\rho\sigma}=-4\alpha\epsilon_{\mu\nu\rho\sigma}\,.
\label{eq_:HAdS}
\ee
Now, $\epsilon_{\mu\nu\rho\sigma}$ is the Levi-Civita tensor of the corresponding space with metric $\gamma_{\mu\nu}$. The configuration of the 3-form should then be such that Eq.~\eqref{eq_:HAdS} holds. Obviously, the configuration for the 3-form is not unique due to its gauge invariance. We do not necessarily require the explicit representation of the 3-form, but we will provide a concrete manifestation. If we consider the following configuration
\be
A_{\mu\nu\rho}=b(x^\alpha)\epsilon_{\mu\nu\rho\lambda}x^\lambda\,,
\label{eq:Bgammab}
\ee
with an arbitrary function on the brane $b(x)$, the field strength is
\be
H_{\mu\nu\rho\sigma}=-\left[\frac{1}{\sqrt{-\gamma}}x^\lambda\partial_\lambda\left(\sqrt{-\gamma} b\right)+4b\right]\epsilon_{\mu\nu\rho\sigma}\,.
\label{eq:Hgammab}
\ee
Therefore, it is straightforward to recover \eqref{eq_:HAdS} if the function $b$ satisfies
\be
\frac{1}{\sqrt{-\gamma}}x^\lambda\partial_\lambda\left(\sqrt{-\gamma} b\right)+4b =4\alpha\,.
\ee 
We will refrain from presenting the intricate details of the construction of the non-trivial realisation of the residual symmetries for curved branes. This process is analogous to the construction outlined above for the flat case but is slightly more tedious. That the residual symmetries will be realised with appropriate choices of the gauge parameters is guaranteed since the physical quantity, i.e., the field strength \eqref{eq_:HAdS}, is invariant. All the expressions will remain identical to those for the Minkowski brane, so ${Y=8\alpha^2 e^{-8W}}$ and ${\Hd^A=4\alpha e^{-4W}\delta^A_y}$ will still yield the same energy densities and pressures. The sole distinction appears in Einstein equations \eqref{field2A}, which now include a term due to the curvature of the brane:
\ba
3\left(W''+2W'^2\right)-\Lambda e^{-2W} &=& F-2Y F_Y\,, \\
6W'^2-2\Lambda e^{-2W} &=& F\,.
\ea
From these equations we have
\be
3W''=-\Big(2YF_Y+\Lambda e^{-2W}\Big)\,,
\ee
which generalizes the Minkowski case presented in Sec.~\ref{sec::brane_eqs}.
We can now compute the slow warping parameters:
\be
\varepsilon_1=-\frac{W''}{W'^2}=\frac{2\Lambda e^{-2W}+4YF_Y}{2\Lambda e^{-2W}+F}
\ee
and
\be
\varepsilon_2=2\varepsilon_1-2\frac{\Lambda e^{-2W}+8Y(F_Y+YF_{YY})}{\Lambda e^{-2W}+2YF_Y}\,,
\ee
while the propagation speed can be expressed as
\be
c_y^2=-\frac{1}{1+\frac14\left[\left(1+\frac{\Lambda e^{-2W}}{YF_Y}\right)(\varepsilon_2-2\varepsilon_1)+\frac{\Lambda e^{-2W}}{YF_Y}\right]}\,.
\ee
We observe that, in the regime where ${\varepsilon_{1,2}\ll1}$, only an Anti-de Sitter brane could potentially be free from gradient instabilities, and this would only be achievable provided ${-\frac{\Lambda e^{-2W}}{YF_Y}>1}$. Nevertheless, this would contradict a scenario in which ${W''<0}$.

As a final remark, let us notice that the field strength \eqref{eq:Hgammab} derived from the 3-form configuration \eqref{eq:Hgammab} does not rely on the metric $\gamma_{\mu\nu}$ being maximally symmetric so, in particular, it will also be valid to describe cosmological thick branes. The residual symmetries on the cosmological brane will be fewer than in the maximally symmetric case, so we could relax the symmetries requirements for the 3-form as to only realise ISO(3), SO(4) or SO(3,1) for flat, closed and open cosmologies respectively. In the language of the scalar dual, the cosmological branes require a scalar field background of the form $\phi=\phi(t,y)$. In these scenarios, the propagation speeds will acquire a dependence on the time-profile of the background field and one might hope for this to cure the unveiled gradient instabilities for the maximally symmetric branes. This might not seem to be a very promising route for relatively standard scenarios since the modes with frequencies larger than the time-variation scale of the brane Hubble factor should be insensitive to the expansion and, consequently, would be prone to instabilities. Let us recall again that the obtained instability is a UV instability. To be a bit more precise, imposing a negative determinant of the effective metric (so there is a well-defined light cone for the perturbations as discusses above)) leads to the necessary (although not sufficient) condition $F_Y+2YF_{YY}>0$ or, equivalently, to $P_{X}+2XP_{XX}>0$ in the scalar dual, where now $Y$ and $X$ include the time dependence due to the cosmological evolution of the brane. If the cosmological evolution is confined to the brane's position so it approaches a maximally symmetric configuration in the asymptotic regions, we effectively have that the $y$-gradients dominate and the time-dependence will only give contributions that will be suppressed by the cosmological slow roll parameters. Thus, our expressions obtained above will be correct up to cosmological slow roll corrections which, being small, cannot cure the gradient instabilities. It would be interesting to study how a generic cosmological evolution affects our findings to confirm our reasoning to show explicitly that the gradient instabilities persist or to explore possible caveats in the outlined argument. We will not pursue this path here. Instead, we will now briefly discuss the case of massive 3-forms where the gauge symmetry is no longer available.

\section{Massive 3-forms}
\label{sec:Massive}

It is evident that the dualisation to a shift-symmetric scalar field fails if the 3-form is massive. Instead, the dualisation of a massive 3-form in 5 dimensions leads to a massive 1-form. This is usually referred to as spin jump, and it simply reflects the different number of degrees of freedom between the massive and massless forms. Let us see how this comes about. Again, we start from the first order formulation of a 3-form
\be
\lag{}=\Pi^{ABCD}H_{ABCD}-\mathcal{H}(\Pi,A)\,,
\ee
with ${H_{ABCD}=4\partial_{[A} A_{BCD]}}$. As usual, we can solve the algebraic momentum equation
\be
H_{ABCD}=\frac{\partial\mathcal{H}}{\partial\Pi^{ABCD}}
\label{eq:PieqToH}
\ee
to obtain $\Pi_{ABCD}$ as a function of $H_{ABCD}$. Then we integrate it out and find the second order formulation of the theory with ${\lag{}=\lag{}(H,A)}$. Alternatively, we can express the above Lagrangian in terms of the duals of $\Pi$ and $A$ to obtain
\be
\lag{\rm dual}=\tilde{\Pi}^{AB}\partial_{[A} \tilde{A}_{B]}-\mathcal{H}(\tilde{\Pi},\tilde{A})\,.
\ee
This shows the advertised equivalence between a massive 3-form and a massive 1-form. We could now integrate out the 1-form conjugate momentum and obtain the corresponding second order formulation of the theory. 

At this point, the equivalence poses a curious riddle. From the perspective of the 1-form, we can take the configuration ${A_M=A(y)\delta^y_M}$, with some function $A$, that respects the desired residual symmetries for the brane in a similar manner to a scalar field with a gradient orthogonal to the brane. This would lead us to think that its dual 3-form description should also allow for a configuration that respects the residual symmetries on the brane. However, one quickly realises that there is no configuration for the 3-form that trivially\footnote{Let us note that the massive 3-form does not have any internal symmetries that we can use to realise the symmetries in a non-trivial way as we did for the massless 3-form. Thus, a massive 3-form must trivially realise the symmetries.} realises the brane symmetries because any non-trivial component for $A_{y\mu\nu}$ breaks the 4-dimensional translations (from the brane perspective, this is just a two form which cannot realise the residual symmetries) and, similarly, for any non-trivial components $A_{\mu\nu\rho}$ (there is no Lorentz invariant object with these indices). Thus, from the 3-form perspective it looks like it is not possible to construct a configuration giving rise to the desired residual symmetries. The resolution to this paradoxical result comes from the (non-)equivalence of the first and the second order formulations of both dual theories. 

When examining the dualisation between the first order formulations of the theories outlined above, there appears to be no obstruction, as it mainly involves introducing the corresponding dual fields. In these formulations, the 1-form configuration has ${A_M=A(y)\delta^y_M}$ and ${\tilde{\Pi}^{AB}=0}$ since a 2-form cannot trivially realise the symmetries and, hence, it must trivialise. The dual description will then have ${A_{MNP}=0}$ and ${\Pi^{ABCD}=\epsilon^{ABCDy}A(y)}$, i.e., its only non-vanishing components will be ${\Pi^{\mu\nu\rho\sigma}=A(y)\epsilon^{\mu\nu\rho\sigma}}$ that respects the symmetries of the brane. Thus, we have reached the conclusion that we must have a non-trivial momentum for an identically vanishing 3-form\footnote{Recall that the dualisation essentially corresponds to a canonical transformation where momenta and coordinates are exchanged.} which is at odds with a second order formulation. The resolution to this conundrum comes from realising that when transitioning from the first order to the second order formulations, certain non-degeneracy conditions must be fulfilled. In order to show the problem let us assume that ${\mathcal{H}=\mathcal{H}(\Pi^2,A^2)}$. Then, the square of the momentum equation \eqref{eq:PieqToH} reads
\be
H^2=4\left(\frac{\partial\mathcal{H}}{\partial\Pi^2}\right)^2\Pi^2
\label{eq:PieqToHsq}
\ee
and we can, in principle, obtain ${\Pi^2=\Pi^2(H^2,A^2)}$. This relation should be continuous and differentiable for the passage to the second order formulation to be well-defined. If we take the derivative with respect to $\Pi^2$ of the right-hand side of Eq.\eqref{eq:PieqToHsq} we obtain the following non-degeneracy condition:
\be
\frac{\partial\mathcal{H}}{\partial\Pi^2}\left(\frac{\partial\mathcal{H}}{\partial\Pi^2}+2 \frac{\partial^2\mathcal{H}}{\partial\Pi^2\partial\Pi^2}\Pi^2\right)\neq0\,,
\ee
that guarantees differentiability of ${\Pi^2=\Pi^2(H^2,A^2)}$. Now, we need to note that, if the 3-form vanishes identically, then its field strength also vanishes identically and Eq.\eqref{eq:PieqToHsq} tells us that ${\frac{\partial\mathcal{H}}{\partial\Pi^2}=0}$ if $\Pi^2$ is to be different from zero. However, this violates the above non-degeneracy condition and, therefore, the Legendre transformation that relates both formulations is ill-defined. It is possible to understand the same obstruction from the perspective of the 1-form. Since $\tilde{A}$ is the dual of the 3-form conjugate momentum $\Pi$, we see that ${\frac{\partial\mathcal{H}}{\partial\Pi^2}=0}$ is equivalent to having ${\frac{\partial\mathcal{H}}{\partial A^2}=0}$. Now, the equation for the dual 1-form for the configuration ${A_M=A(y)\delta^y_M}$ and ${\tilde{\Pi}^{AB}=0}$ is simply ${\frac{\partial\mathcal{H}}{\partial A^2}A(y)=0}$. Thus, the non-trivial solution ${A(y)\neq0}$ requires ${\frac{\partial\mathcal{H}}{\partial A^2}=0}$, which precisely coincides with a singular point of the Legendre transformation. This resolves the riddle and there is no contradiction in having a 1-form realising the brane symmetries while a 3-form cannot achieve it due to the failure of the dualization of the two second order formulations.\footnote{A similar situation occurs for cosmological models with massive 2-forms and their 1-form duals \cite{Aoki:2022ylc}.}

The above argument does not imply a strict no-go result for the construction of branes using 3-forms. Rather, it suggests that non-standard approaches or additional requirements will be necessary for such realisations. For instance, one might find the brane construction satisfactory using the first order formulation even if the second order formulation is not feasible considering the caveats that come with it. On the other hand, a thick brane realisation with massive 3-forms has been recently suggested in \cite{Gordin:2023nsv}. The trick in that realisation is that the residual symmetries on the brane are not realised off-shell by the 3-form field. Instead, an on-shell condition needs to be imposed to be compatible with the brane symmetries, giving rise to additional requirements.

\section{Conclusions}\label{sec:conclusions}

In this work we have explored non-trivial realisations of thick branes with 5-dimensional massless 3-forms. We have commenced by considering Minkowskian branes and we have explicitly shown how the action of the 4-dimensional Poincar\'e group on the 3-form can be compensated by an internal gauge transformation through a suitable choice of the gauge parameter. In this manner, the residual Poincar\'e symmetry is non-trivially realised by the combined action of spacetime and gauge transformations. We have then studied the 5-dimensional Einstein equations for the construction of the thick brane solutions and obtained some viability conditions for the absence of pathologies for the 3-form. An important result of our analysis is that there is an incompatibility between having a brane profile that decays slowly in the asymptotic region far from the brane position and the absence of gradient instabilities. In particular, our results exclude the most common brane profiles in the literature which, hence, cannot be constructed with massless 3-forms. 

After analysing the viability of the thick brane solutions supported by 3-form fields, we have resorted to the dualisation of the 3-form to a shift-symmetric scalar field. We have explicitly constructed the dualisation and obtained the on-shell relations between both formulations. We have corroborated our findings from the scalar field perspective and, once again, demonstrated that slowly decaying brane profiles in the asymptotic region necessary incur gradient instabilities. We have illustrated our results with an explicit example. Although our results have been obtained for shift-symmetric scalar fields, they can be applied to non-shift-symmetric theories when the scalar field evolution exhibits an approximate shift-symmetry in the asymptotic regions. 

In most of the work, we have considered a Minkowskian (flat) brane. However, we have also analysed the case of non-flat AdS/dS branes. We have constructed the 3-form configuration that realises the corresponding symmetries (modulo a gauge transformation) and obtained the Einstein equations for the curved branes. Although the curvature of the brane might alleviate the pathological nature of the brane profile, we have shown that this is not the case for a dS brane, while the AdS brane would require a too large value for the cosmological constant. Hence, adding curvature to the brane does not resolve the problems found for the Minkowskian branes.

Finally we have briefly commented on the case of massive 3-forms. The dualisation of these theories takes us to a massive 1-form and we have discussed the paradoxical result that a 1-form can trivially realise the brane symmetry while a 3-form cannot. However, we have diagnosed the apparent contradiction being caused by the failure of the dualisation for the required 1-form configuration.

In this work we have mainly focused on one single 3-form field. There are however other possibilities to non-trivially realise the residual symmetries on the brane. For instance, we could take a set of four 3-forms ${A^a{}_{MNP}}$ where $a$ is an internal Lorentz index. Thus, we could construct the configurations ${A^a_{\mu\nu\rho}\propto \epsilon^a{}_{\mu\nu\rho}}$ so that spacetime Lorentz transformations combined with the internal ones would leave a diagonal Lorentz group invariant. Another possibility would be to adopt four 2-form fields $A^a_{MN}$ with an internal global SO(3,1) symmetry so a background configuration of the form ${A^a_{\mu\nu}=\epsilon^a{}_{\mu\nu\rho} x^\rho}$ would respect a diagonal 4-dimensional Lorentz group. It is worth noticing that having an internal symmetry corresponding to a non-compact group might make more contrived to construct explicit models since the corresponding Killing metric is non-positive definite with the associated ghostly modes. There exists an extensive zoology of brane constructions and their corresponding dual relations, providing an interesting avenue for exploration that we reserve for future research.

\acknowledgments
B.J.B. acknowledges support from the Funda\c{c}\~{a}o para a Ci\^{e}ncia e a Tecnologia (FCT) through the project BEYLA: BEYond LAmbda, with reference number PTDC/FIS-AST/0054/2021.  J.B.J. was supported by the Project PID2021-122938NB-I00 funded by the Spanish “Ministerio de Ciencia e Innovaci\'on" and FEDER “A way of making Europe”.

\appendix
\section{Non-flat branes}
In this Appendix we give the 3-form configurations for the non-flat branes in some common coordinate systems. In all the cases, the 3-form is constructed so that ${H_{\mu\nu\rho\sigma}=-4\alpha \epsilon_{\mu\nu\rho\sigma}}$ and $\epsilon_{\mu\nu\rho\sigma}$ is the corresponding Levi-Civita tensor. The 3-form configuration is ${B_{\mu\nu\rho}=b(x^\mu)\epsilon_{\mu\nu\rho\sigma}}$, as explained in the main text, and the equation satisfied by $b(x)$ depends on the coordinate system as we detail in the following:

\begin{itemize}
\item Conformal coordinates. In these coordinates, the (A)dS metrics can be written as
\ba
\dd s^2_{\text{dS}}&=&\frac{e^{2W}}{H^2t^2}\left(-\dd t^2+\dd \vec{x}^2\right)+\dd y^2\,,\\
\dd s^2_{\text{AdS}}&=&\frac{e^{2W}}{H^2x_3^2}\left(-\dd t^2+\dd x_1^2+\dd x_2^3+\dd x_3^2\right)+\dd y^2\,,
\ea
and the equation for $b$ adopts a particularly simple form
\be
x^\mu\partial_\mu b=4\alpha\,.
\ee
\item Planar coordinates. The line elements are given by
\ba
\dd s^2_{\text{dS}}&=&e^{2W}\left(-\dd t^2+e^{2Ht}\dd \vec{x}^2\right)+\dd y^2\,,\\
\dd s^2_{\text{AdS}}&=&e^{2W}\left[e^{2Hx_3}\Big(-\dd t^2+\dd x_1^2+\dd x_2^3\Big)+\dd x_3^2\right]+\dd y^2\,,
\ea
and the equations satisfied by $b(x)$ are
\begin{align}
&\text{dS:}\qquad\Big(x^\mu\partial_\mu +4+3Ht\Big)b=4\alpha\,,\\
&\text{AdS:}\quad\Big(x^\mu\partial_\mu +4+3Hx_3\Big)b=4\alpha\,,
\end{align}
respectively. The Minkowski brane is then recovered by sending $H\to0$ so we have the solution ${b=\alpha}$.

\item Static coordinates. In these coordinates we have
\be
\dd s^2_{\text{Brane}}=-\left(1-\frac13\Lambda r^2\right)\dd t^2+\frac{1}{1-\frac13\Lambda r^2}\dd r^2+r^2\dd \Omega^2\,,
\ee
for both dS ($\Lambda>0$) and AdS ($\Lambda<0$) as well as the flat brane with $\Lambda=0$. The equation for $b$ is now
\be
\left(x^\mu\partial_\mu +6+\frac{\theta}{\tan\theta}\right)b=4\alpha\,.
\ee
It is interesting to note that the 3-form background configuration does not depend on the curvature $\Lambda$ in these coordinates. 
\end{itemize}

\section{Dual relations}\label{appendix}
In this Appendix, we will provide the detailed derivation of the dualisation of the massless 3-form in 5 dimensions to a shift-symmetric scalar field. We will introduce the object ${\epsilon_{ABCDE}=\sqrt{-g}\,\varepsilon_{ABCDE}}$ denotes the (coordinate-invariant) Levi-Civita symbol. Let us begin by considering a first order formulation:
{
\be\label{first_order_action}
\mathcal{S}=\int \dd^5x\sqrt{-g}\left[ F(Y)-\frac{1}{24}\frac{\partial F}{\partial Y}P^{ABCD}\left(H_{ABCD}-P_{ABCD}\right) \right]\,,
\ee}
with
\be
Y = -\frac{1}{48}P_{ABCD}P^{ABCD}\,,
\ee
and where $P^{ABCD}$ is an auxiliary 4-form field. We can define the Hodge dual
\be
\tilde{P}^A:=\star P=\frac{1}{24}\epsilon^{BCDEA}P_{BCDE}\,,
\ee
which can be inverted to give
\be
P_{ABCD}=\epsilon_{ABCDE}\tilde{P}^E\,.
\ee
Plugging into the above action gives
\be
\mathcal{S}=\int \dd^5x\sqrt{-g}\left[ F(Y)+\frac{\partial F}{\partial Y}\tilde{P}_A\left(\tilde{H}^A-\tilde{P}^A\right) \right]\,,
\ee
where
\be
Y = \frac{1}{2}\tilde{P}_A\tilde{P}^A\,,
\ee
noting that ${\epsilon_{ABCDE}\epsilon^{ABCDF}=-24\delta^F_E}$. Variation with respect to ${\tilde{P}^{A}}$ gives
\be
\left(\tilde{H}^A-\tilde{P}^A\right)\left(\frac{\partial F}{\partial Y}g_{AB}+\frac{\partial^2F}{\partial Y^2}\tilde{P}_A\tilde{P}_B\right)=0\,,
\ee
which yields the solution ${\tilde{H}^A=\tilde{P}^A}$ for
\be
{\text{det}}\left(\frac{\partial F}{\partial Y}g_{AB}+\frac{\partial^2F}{\partial Y^2}\tilde{P}_A\tilde{P}_B\right)\neq 0\,.
\ee

Defining now
\be\label{def1}
\Pi^{ABCD}=\frac{\partial F}{\partial Y}P^{ABCD}\,,
\ee
we can rewrite the action Eq.~\eqref{first_order_action} as,
\be
\mathcal{S}=\int\dd^5x\sqrt{-g}\left[-\frac{1}{6}\Pi^{ABCD}\partial_{[A}A_{BCD]}+\mathcal{H}(Y_{\Pi})\right]\,,
\ee
with
\be
Y_{\Pi}=-\frac{1}{48}\Pi_{ABCD}\Pi^{ABCD}\,,
\ee
and
\be\label{relation_P_F}
\mathcal{H}(Y_{\Pi})=\left[F(Y_P)-2Y_P\frac{\partial F}{\partial Y_P}\right]_{Y_P=Y_P(Y_{\Pi})}\,.
\ee
Remembering the definition for the strenght tensor for the 3-form, i.e., ${H_{ABCD}=4\partial_{[A}A_{BCD]}}$, the equations of motion for $\Pi$ are found to be
\be
H_{ABCD}=-\frac{\partial \mathcal{H}}{\partial Y_{\Pi}}\Pi_{ABCD}\,.
\ee
This equation gives the explicit relation between the field strength {\bf H} and the conjugate momentum {\bf $\Pi$}. The equation of motion for the 3-form, ${\nabla\cdot\Pi=0}$, is solved for,
\be\label{sol1}
\Pi_{ABCD}=\epsilon_{ABCDE}\partial^{E}\phi\,,
\ee
from which, by squaring the above relation, we get ${Y_{\Pi}=-X}$, and we arrive at the following dual action in terms of the scalar $\phi$,
\be
\mathcal{S}_{\rm dual}=\int\dd^5x\sqrt{-g}\mathcal{H}(-X)\equiv\int\dd^5x\sqrt{-g}\mathcal{P}(X)\,.
\ee

The following relations hold on-shell,
\ba
H_{ABCD}&=&\frac{\partial \mathcal{P}}{\partial X}\epsilon_{ABCDE}\partial^E\phi\,,\label{rel1}\\
Y&=& -X\left(\frac{\partial \mathcal{P}}{\partial X}\right)^2\label{aux}\,.
\ea
and Eqs.~\eqref{def1}, \eqref{sol1} toghether with $\tilde{P}^A=\tilde{H}^A$, gives
\be\label{uno}
\frac{\partial F}{\partial Y}H_{ABCD}=\epsilon_{ABCDE}\partial^E\phi\quad\Rightarrow\quad X=-Y\left(\frac{\partial F}{\partial Y}\right)^2\,.
\ee

We can express Eq.~\eqref{rel1} as
\be \frac{\partial \mathcal{P}}{\partial X}\partial_E\phi = -\frac{1}{24}\epsilon_{ABCDE}H^{ABCD}\,, \ee
which together with Eq.~\eqref{uno} read,
\ba
\star\left(\frac{\partial \mathcal{P}}{\partial X}\dd\phi\right)&=&-\dd {\bf A} \,,\\
\star\left(\frac{\partial F}{\partial Y}\dd {\bf A}\right) &=& \dd \phi\,.
\ea

Equation \eqref{relation_P_F} allows us to write the relation between a given $\mathcal{P}(X)$ and an $F(Y)$ theory,
\be \mathcal{P}=F-2Y\frac{\partial F}{\partial Y}\,, \ee
where $Y\equiv Y(X)$. Analogously, the inverse transformation can be obtained by following the same procedure starting from the first order formalism
\be
\mathcal{S}=\int \dd^5 x \sqrt{-g}\left[\mathcal{P}(X) - \frac{\partial \mathcal{P}}{\partial X}p^{A}\left( 
\partial_A\phi - p_A \right)\right]\,,
\ee
where now ${X=-p_Ap^A/2}$ with $p_A$ an auxiliary vector field. Integrating out this vector, we then identify
\be
\frac{\partial \mathcal{P}}{\partial X}p^A = \frac{1}{24}\epsilon^{ABCDE}\partial_{[B}A_{CDE]}\,,
\ee
and establish the relation,
\be
F = \mathcal{P} - 2X \frac{\partial \mathcal{P}}{\partial X}\,.
\ee

Thus, we can finally summarize the dual relations between $\mathcal{P}(X)$ and $F(Y)$ as
\ba
\mathcal{P} &=& F - 2Y \frac{\partial F}{\partial Y}\,,\quad\quad \star\left(\frac{\partial F}{\partial Y}\dd {\bf A}\right) = \dd \phi\,, \\
F &=& \mathcal{P}-2X\frac{\partial\mathcal{P}}{\partial X}\,,\quad\quad\star\left(\frac{\partial \mathcal{P}}{\partial X}\dd\phi\right)=-\dd {\bf A}\,,\quad
\ea
Using the above relations it is straightforward to find:
\ba
\dd Y &=& -\frac{\partial \mathcal{P}}{\partial X}\left( \frac{\partial \mathcal{P}}{\partial X} + 2X\frac{\partial^2 \mathcal{P}}{\partial X^2} \right)\dd X\label{dual31}\,,\\
\dd X &=& -\frac{\partial F}{\partial Y}\left( \frac{\partial F}{\partial Y} + 2Y\frac{\partial^2 F}{\partial Y^2} \right)\dd Y\,,
\ea
where we have the relations:
\be
\frac{\partial F}{\partial Y}=\left(\frac{\partial \mathcal{P}}{\partial X}\right)^{-1}\,,
\ee
and
\be
\frac{\partial F}{\partial Y} + 2Y\frac{\partial^2 F}{\partial Y^2}= \left( \frac{\partial \mathcal{P}}{\partial X} + 2X\frac{\partial^2 \mathcal{P}}{\partial X^2} \right)^{-1}\,.
\ee
The denominators on the last two relations are required to be non-singular in order to have a well-defined passage to the first order formalism employed to derive them.

\bibliography{bib}

\end{document}